\documentclass[]{spie}   
\usepackage{graphicx}
\usepackage{subfig,amsmath}
\usepackage{caption}

\usepackage[suffix=]{epstopdf}
\usepackage{hhline}
\usepackage{booktabs}
\usepackage[table,xcdraw]{xcolor}

\newcommand{\fig}[1]{Fig.~\ref{#1}}

\usepackage{color}

\title{Commissioning and performance results of the WFIRST/PISCES integral field spectrograph}
\author{Prabal Saxena\supit{a}, Maxime J. Rizzo\supit{a}, Camilo Mejia Prada\supit{b}, Jorge Llop Sayson\supit{c}, Qian Gong\supit{a}, Eric J. Cady\supit{b}, Avi M. Mandell\supit{a}, Tyler D. Groff\supit{a}, Michael W. McElwain\supit{a}
\skiplinehalf
\supit{a}Goddard Space Flight Center, 8800 Greenbelt Road, Greenbelt, MD, 20771, USA; \\
\supit{b}Jet Propulsion Laboratory, California Institute of Technology, 4800 Oak Grove Drive, Pasadena, CA, 91109 USA;
\\
\supit{c}California Institute of Technology, 1200 E. California Boulevard, Pasadena, CA, USA
}

\authorinfo{Send correspondence to Prabal Saxena \texttt{prabal.saxena@nasa.gov}}

\begin{document}
\maketitle
\begin{abstract} 
The Prototype Imaging Spectrograph for Coronagraphic Exoplanet Studies (PISCES) is a high contrast integral field spectrograph (IFS) whose design was driven by WFIRST coronagraph instrument requirements.  We present commissioning and operational results using PISCES as a camera on the High Contrast Imaging Testbed at JPL.  PISCES has demonstrated ability to achieve high contrast spectral retrieval with flight-like data reduction and analysis techniques.    
\end{abstract}
\section{Introduction}\label{sec:intro}

Direct imaging of exoplanets has become a priority in the field of exoplanet discovery and characterization due to its ability to directly obtain evidence about a planet's atmosphere and some bulk properties.  Features such as atmospheric composition, structure and clouds are just some of the planetary properties obtainable from directly imaged spectra.  However, detecting and observing spectra of exoplanets using direct imaging is challenging due to the combination of extreme star to planet contrast ratios and the relatively small apparent physical separation between a host star and an orbiting planet.  Detection of Earth-sized planets in reflected visible light requires contrast ratios of $10^{10}$, while even detection of Jupiter-sized planets and large young self-luminous planets requires contrast ratios of $10^{8}$ and $10^{6}$, respectively.  Consequently, direct detection of exoplanets requires observing strategies which push the boundaries of high contrast imaging.  The use of coronagraphy to occult a host star has been combined with adaptive optics (AO) technology to yield a particularly promising means of potentially achieving the required contrast ratios in regions close-in enough to the host star.  Ground based adaptive optics systems such as The Gemini Planet Imager (GPI) \cite{2006SPIE.6272E..0LM} and Spectro-Polarimetric High-contrast Exoplanet REsearch (SPHERE) \cite{2015A&A...576A.121M} instrument have been able to achieve contrast ratios nearing $10^{7}$ using post-processing techniques \cite{2014PNAS..11112661M, Macintosh64} and have yielded a number of direct detections of young self-luminous planets.  Advancing these technologies onto a space based platform immune to the difficulties posed by the effects of Earth's atmosphere is the next step in accessing even larger contrast ratios.

\subsection{PISCES Integral Field Spectrograph: High Contrast Imaging and WFIRST} 
\label{sec:PISCES}

The Prototype Imaging Spectrograph for Coronagraphic Exoplanet Studies (PISCES) was designed by Michael McElwain to address the need for an Integral Field Spectrograph (IFS) to operate at the high contrast regime targeted by space missions.  PISCES was specifically designed to advance towards the high contrast imaging requirements needed to observe Exo-Earths around nearby stars.  As part of its demonstration of spectral retrieval at high contrast, PISCES sought to advance technology readiness levels for future exoplanet science cameras and the high contrast imaging testbed (HCIT) at the Jet Propulsion Laboratory (JPL), where it would be tested.  Additionally, PISCES was to demonstrate and improve full life cycle operations for an IFS, beginning with wavefront sensing and control algorithms through to realistic post-processing demonstrations.  

However, in 2012 after receiving a 2.4 m telescope from the National Reconnaissance Office (NRO), the Wide Field Infrared Survey Telescope (WFIRST) mission had a Science Definition Team (SDT) chartered by NASA to examine the capabilities the mission could have with such a telescope.  The SDT explored the option of including a coronagraphic instrument that would enable imaging and spectroscopic studies of planets around nearby stars. \cite{2013arXiv1305.5422S}  The WFIRST Coronagraph Instrument (CGI) project requested a revised PISCES design with the same detector format but with different design specifications.  This revised PISCES design is the one described here and represents a viable flight configuration for the WFIRST Coronagraph Instrument (CGI).

\section{PISCES Instrument Design} 

\subsection{Instrument Specifications} 
\label{sec:InstSpec}

Instrument specifications that guided design of PISCES were given by the 2015 WFIRST SDT and are listed in table \ref{instrumentspecs}.

\begin{table}[h!]
\centering
\caption{PISCES instrument specifications.}
\begin{tabular}{|l|c|c|c|}
\hline
\multicolumn{4}{|c|}{\cellcolor[HTML]{000000}{\color[HTML]{FFFFFF} \textbf{WFIRST - CGI Specifications}}}             \\ \hline
\begin{tabular}[c]{@{}l@{}}Band Central Wavelengths (nm)\\ PISCES-R, -I, -Z\end{tabular} & 660      & 770    & 890    \\ \hline
$\lambda$min (nm)                                                                                & 600      & 700    & 810    \\ \hline
$\lambda$max (nm)                                                                                & 720      & 840    & 970    \\ \hline
Instantaneous Bandpass ($\delta\lambda/\lambda$c)                                                           & \multicolumn{3}{c|}{0.18}  \\ \hline
f/\#                                                                                     & \multicolumn{3}{c|}{870}   \\ \hline
Lenslet pitch ($\mu$m)                                                                       & \multicolumn{3}{c|}{174}   \\ \hline
Lenslet sampling at $\lambda$c {[}\# lenslets/($\lambda$c/D){]}                                          & 3.3      & 3.9    & 4.6    \\ \hline
FoV (\# of $\lambda$c/D){[}radius{]}                                                             & 14.5     & 12.4   & 10.7   \\ \hline
Pinhole diameter ($\mu$m)                                                                    & \multicolumn{3}{c|}{25-30} \\ \hline
Lenslet array format                                                                     & \multicolumn{3}{c|}{76x76} \\ \hline
Magnification from lenslet to detector                                                   & \multicolumn{3}{c|}{1:1}   \\ \hline
Spectral Resolving Power ($\lambda/\delta\lambda$)                                                          & \multicolumn{3}{c|}{$70\pm5$}  \\ \hline
\end{tabular}
\label{instrumentspecs}
\end{table}


\subsection{Integral Field Spectrograph Design Contraints}

PISCES utilizes a lenslet based IFS, which is advantageous due to its ability to sample the image plane early with a large number of spatial samples.  This enables high image quality preservation.  The basic design of a lenslet driven IFS begins with a lenslet array at the image plane.  The image plane captures and compresses light from different regions of the image and relays this grid of spots to the remaining optical system of the traditional spectrograph.  During this process, the light is collimated and then dispersed using a prism into spatially separated wavelength dependent strips onto the detector.  In order to maximize resolution and allow for as many dispersed spectra as possible to fit on the detector, the lenslet array was clocked at an angle of 26.565 degrees to the detector.  Additionally, the lenslet pitch required to maximize packing efficiency for the PISCES detector is given in table \ref{instrumentspecs}.  Once the two dimensional spectra are collected on the detector, they are reassembled into three dimensional cubes that have a two dimensional spatial component and a wavelength component.  The camera uses a e2v CCD 47-10-1-353 detector.  The PISCES layout, as well as a future flight IFS for WFIRST, are optimized by a simulation package \cite{Rizzo2017} that tests different designs.

\subsection{Optical Design} 
\label{sec:OptDes}

\begin{figure}
\centering
\includegraphics[width=0.8\textwidth]{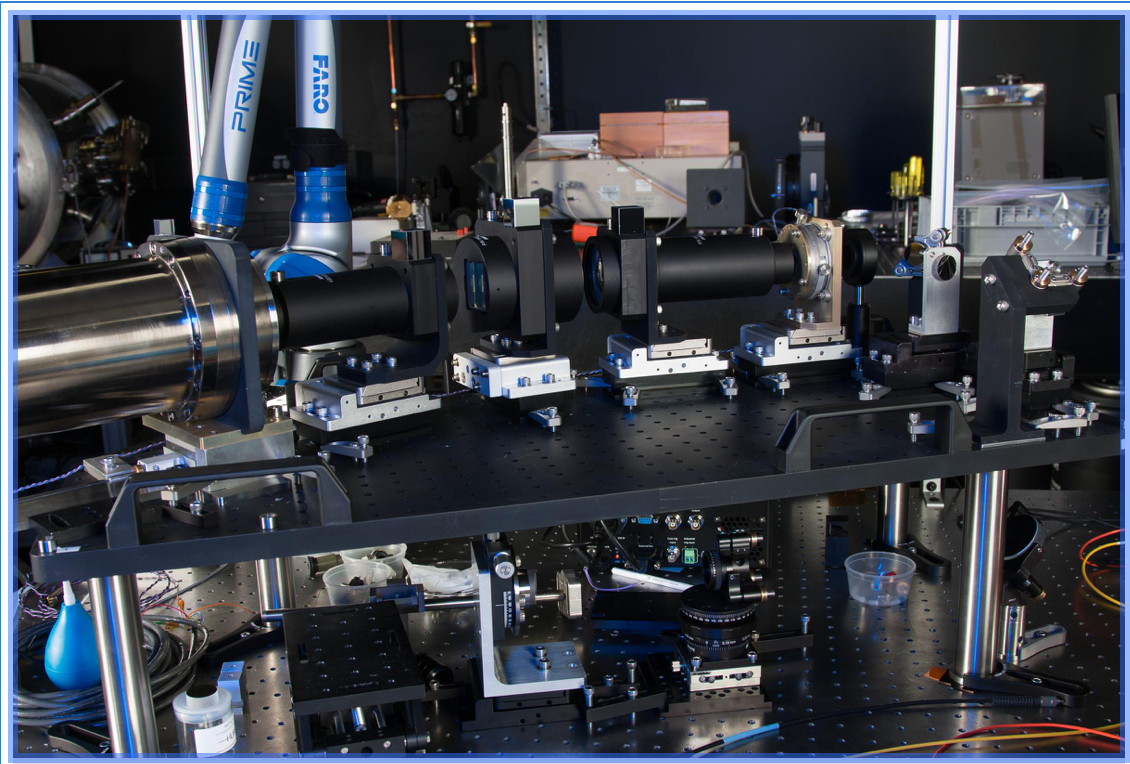}
\caption{An image of assembled optical design of PISCES at the Goddard Integral Field Spectroscopy Laboratory.}
\label{fig:piscesgoddard}
\end{figure}

The optical design for PISCES is described in detail in previous work\cite{gong2015prototype}.  The optics consist of three sub-components - the lenslet array, relay optics and the prism spectrometer. The relay optics and spectrometer design largely follow typical designs with a few notable exceptions that are a consequence of the inclusion of the lenslet array.  The relay adjusts the plate scale to match the desired field of view for the lenslet size.  The spectrometer includes a collimator, compound prism assembly and imager, but also accounts for the transaction of the beam at the lenslet array.  The beam has a new f number of 870 after the lenslet array that was accommodated by the relay optics.  PISCES uses an off-axis Ritchey-Chrétien telescope that provides the magnification necessary and that is fit into the space requirements using two fold mirrors. An image of the fully assembled PISCES instrument at Goddard Space Flight Center is given in figure \ref{fig:piscesgoddard}.

\subsection{Testbed Integration} 
\label{sec:testbed} 

\begin{figure}
\centering
\includegraphics[width=0.8\textwidth]{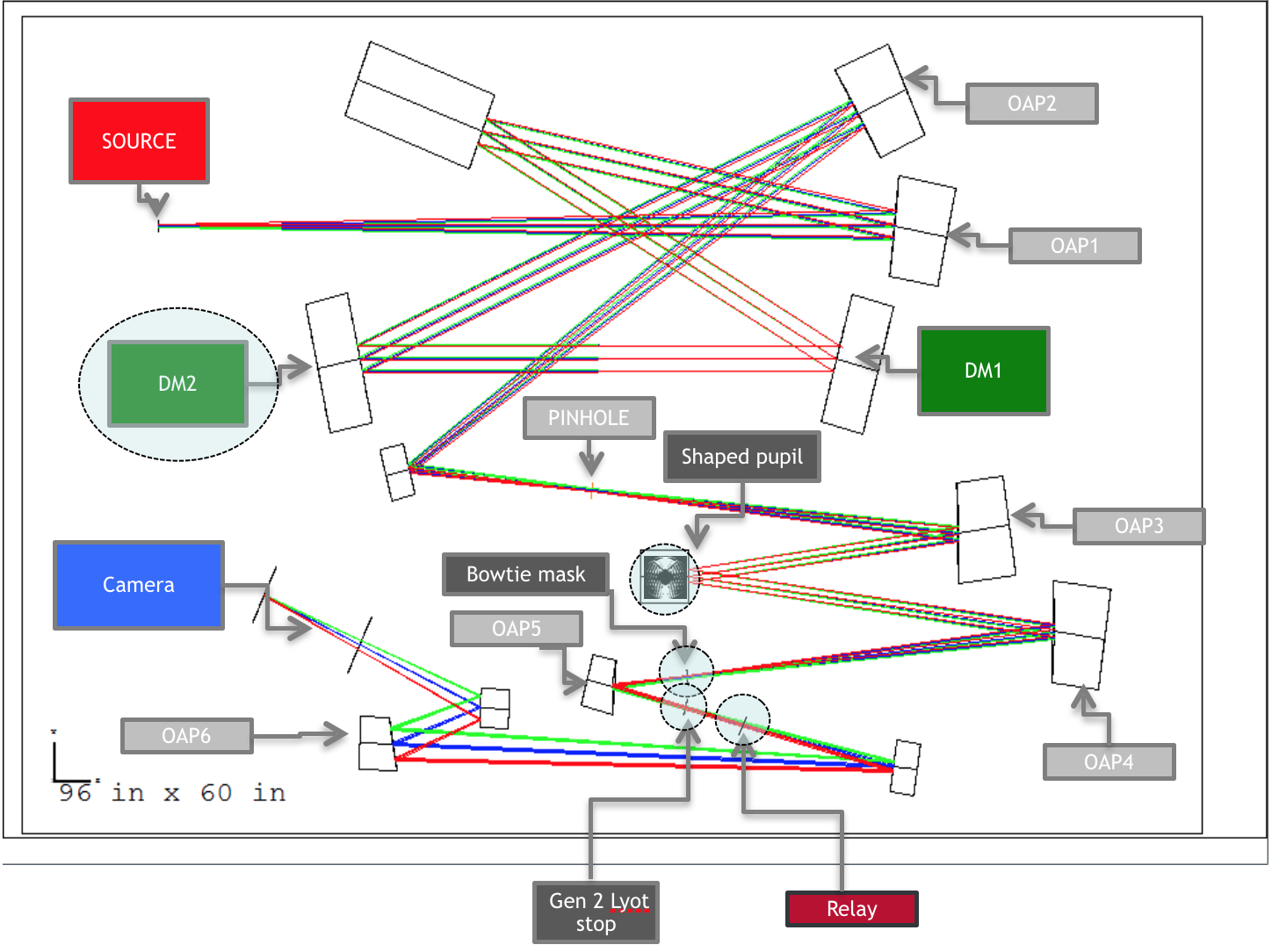}
\caption{The layout of the Shaped Pupil Coronograph testbed at the Jet Propulsion Laboratory.}
\label{fig:piscesspc}
\end{figure}

PISCES was designed so that it could be integrated with three different testbeds at JPL: the Shaped Pupil Coronagraph (SPC) testbed, the Hybrid Lyot Coronagraph (HLC) testbed, and the Occulting Mask Coronagraph (OMC) testbed.  The OMC testbed allows for a dynamic stellar input (versus the static stellar input of the other testbeds), in order to test the ability to remove telescope jitter (results of which are given in corresponding forthcoming papers that include greater description of the testbeds).  The results presented in this paper reflect performance of PISCES in the SPC testbed.  The layout of the SPC testbed is given in figure \ref{fig:piscesspc}. 

\section{Instrument Performance} \label{sec:sections}

\subsection{Calibration and Testing} 
\label{sec:CalibTest}

The optical design of PISCES was tested to evaluate lenslet array performance (including pinhole mask performance), spot size control, spectral dispersion and layout including optical distortion, and background light assessment. The detector was also evaluated by measuring flat fielding, fringing, read noise and bad pixels.  The overall optical throughput of PISCES including the mirror optics, lenslet array, collimator, disperser, and detector was 0.75 (with no individual component being less than 0.95).  Including detector quantum efficiency, PISCES' total efficiency was 0.53.  Detector response peaks at 575nm (and is thus greatest at the shortest wavelength of the bluest band), and is lowest at the reddest wavelengths (with a total efficiency, including optics, of 15$\%$ at 970nm). 

\begin{figure}[b]
\centering
\includegraphics[width=0.8\textwidth]{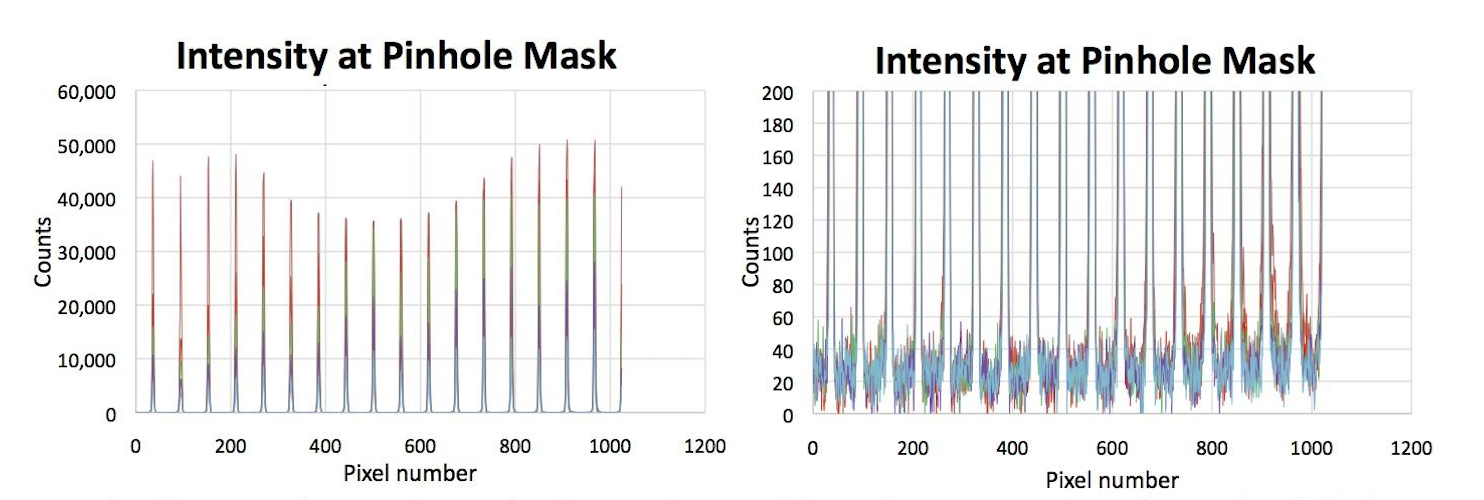}
\caption{Intensity of each lenslet (represented by different colors) at different source flux levels.  The pinhole masks remove excess flux between the PSF core of each lenslet. The right image zooms in on the counts to show the levels of stray light.}
\label{fig:pinholeleak}
\end{figure}

PISCES is designed with a pinhole mask placed at the focus of where a single spatial resolution element of the lenslet array sends light onto a spot behind the array.  This mask is meant to suppress stray diffracted light from neighboring pinholes from contaminating signals for a particular lenslet with crosstalk. PISCES was tested to determine the efficiency of the pinhole mask.  These tests indicated the mask blocks light at $>10^{4}$ level and was demonstrated not to leak light (see figure \fig{fig:pinholeleak}).  Additionally, lenslet intensity maps were created with and without the pinhole mask.  Total flux loss was only 2.5$\%$ with the addition of the mask. 

The designed spot size was tested and verified to ensure that the PSF was nyquist-sampled on the detector.  Additionally, it was also verified that the spot size could meet the desired goal of being undersampled if necessary.  Spot size below 2 detector pixels was desired by the WFIRST CGI team in order to reduce detector noise for what are likely to be faint planetary targets.

Spectra produced by PISCES was aligned to ensure that dispersion was aligned on the detector and to ensure correct spacing between adjacent spectra.  Relative clocking of the lenslet array to the detector ensured there was appropriate space in the direction of cross dispersion between spectra as well as in the direction of the spectral trace.   This clocking also maximized the number of spectra that could fit on the detector.  Clocking of the prism relative to the detector using external metrology ensured correct alignment of the spectra relative to detector rows.  Optical distortion is coupled to spectral dispersion and was also measured.  Grid distortion was measured to be less than 3$\%$, which met design specifications.  Finally, faint source and long exposure observations meant to simulate planetary observations are especially susceptible to contamination by stray light, and precaution was taken to ensure elimination of such sources before and after integration with the HCIT.

After testing optical design specifications, performance of the detector was also evaluated. Flat fielding was measured across the spectral range, with standard deviations for the 633 nm and 775 nm flat field of 0.00406 and 0.00408, respectively.  Pixel to pixel variation between these frames was also small, with a mean of 0.00086 and standard deviation of 0.00375.  Fringing (due to reflection within the ccd) was minimal until wavelengths redder than 810 nm.  The maximum peak to trough amplitude was 80$\%$, which was at 970 nm.  Finally, both read noise and bad pixels were evaluated and were accounted for in future measurements.

\subsection{Operation} 
\label{sec:Oper}

The camera assembly for PISCES operates under vacuum in order for the detector to cool safely.  In the HCIT, the entire PISCES assembly is kept under vacuum ($>$2 Torr) during operation. The detector is controlled externally using a Spectral Instruments GUI and can also be controlled via python command line prompts.  The Spectral Instruments detector provides a LabVIEW VI program to accept commands to control the camera.  Focus adjustment (to control sampling on the detector) and detector and instrument re-alignment were done manually.

The input sources used in conjunction with PISCES were the Laser Line Tunable Filter (LLTF), a 637nm laser and the NKT Photonics EXTREME (EXB-6) super-continuum source which was used with the VARIA tunable filter.  The LLTF and EXB-6 were mainly distinguished by a tradeoff between precision and narrowness in spectral bandwidth and overall throughput.  Wavelength accuracy was especially important during calibration procedures, as it affected contrast measurements in individual wavelength slices in the three dimensional cube.  Both the LLTF and VARIA were also controlled by python scripts. During broadband operation, the NKT/VARIA was used to create an 18\% input spectral band.

\subsection{Calibration and Data Reduction} 
\label{sec:datared}

\begin{figure}
\centering

\subfloat{\includegraphics[width=0.5\textwidth]{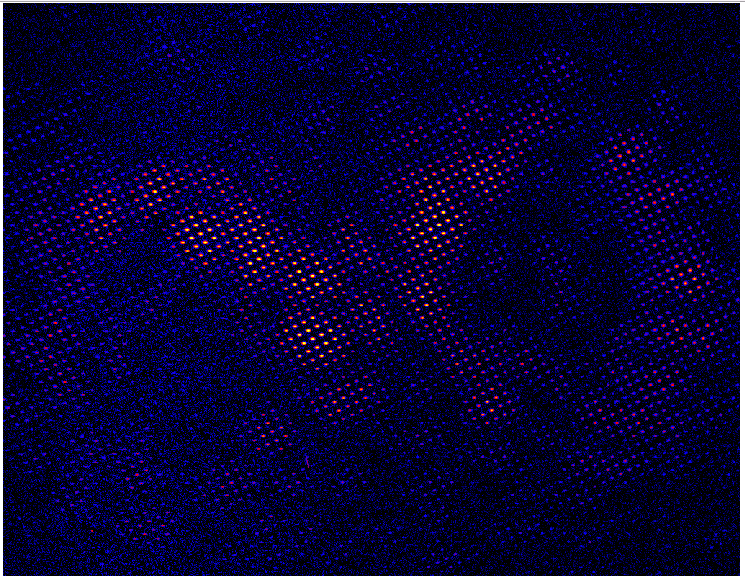}\label{signal_model}}
\subfloat{\includegraphics[width=0.5\textwidth]{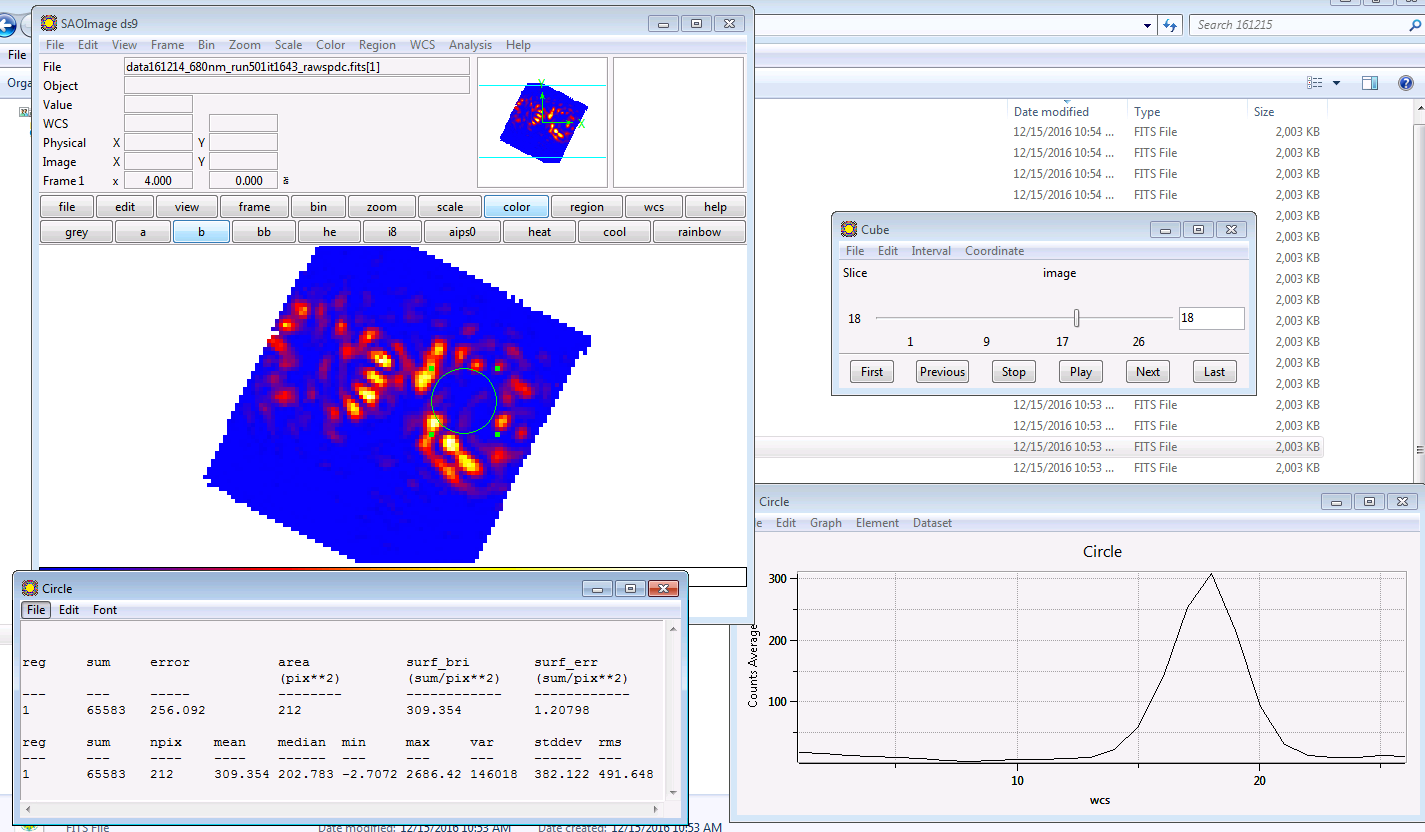}\label{amplitude_mod}}

\caption{Images of a narrow band dark hole at 680 nm: \protect\subref{signal_model} The two dimensional image on the detector. 
  \protect\subref{amplitude_mod} The slice in the three dimensional cube corresponding to the interval containing most of the flux.}

\end{figure}

In order to extract two dimensional data on the detector and convert it to three dimensional wavelength dependent cubes, generation of global wavelength calibration maps were required.  The calibration procedure begins by finding the centroids of pinhole images on the detector - these pinhole images correspond to narrow band images taken within specified uniform wavelength intervals. These pinhole images are centroided by fitting a Gaussian to the data as a function of their x and y position on the detector and are assigned to the appropriate wavelength. These x and y centroid positions are then fit to two sets of polynomials to describe their relationship in the detector plane as a function of wavelength and field position.  The polynomial coefficients derived from the measurements describe the distortion and spectral dispersion.  The size of the wavelength intervals determine the precision of the wavelength calibration as smaller intervals (a typical interval is 10 nm) will ensure more accurate wavelength maps on the detector.    These narrow band centroid maps are then combined to produce a global wavelength calibration that can be used during data reduction.

The data reduction pipeline is built off an open source version of the Gemini$/$GPI pipeline.  The pipeline allows the choice of a combination of different algorithms (termed 'primitives') to form reduction procedures ('recipes').  The default recipe includes subtraction of a dark background if a dark background FITS file is provided and then a flat-field correction of the image.  Finally, the recipe interpolates bad pixels if a bad pixel map is provided.  At this point the recipe can make use of a loaded wavelength calibration to proceed with a spectral extraction.

\begin{figure}
\centering
\includegraphics[width=0.8\textwidth]{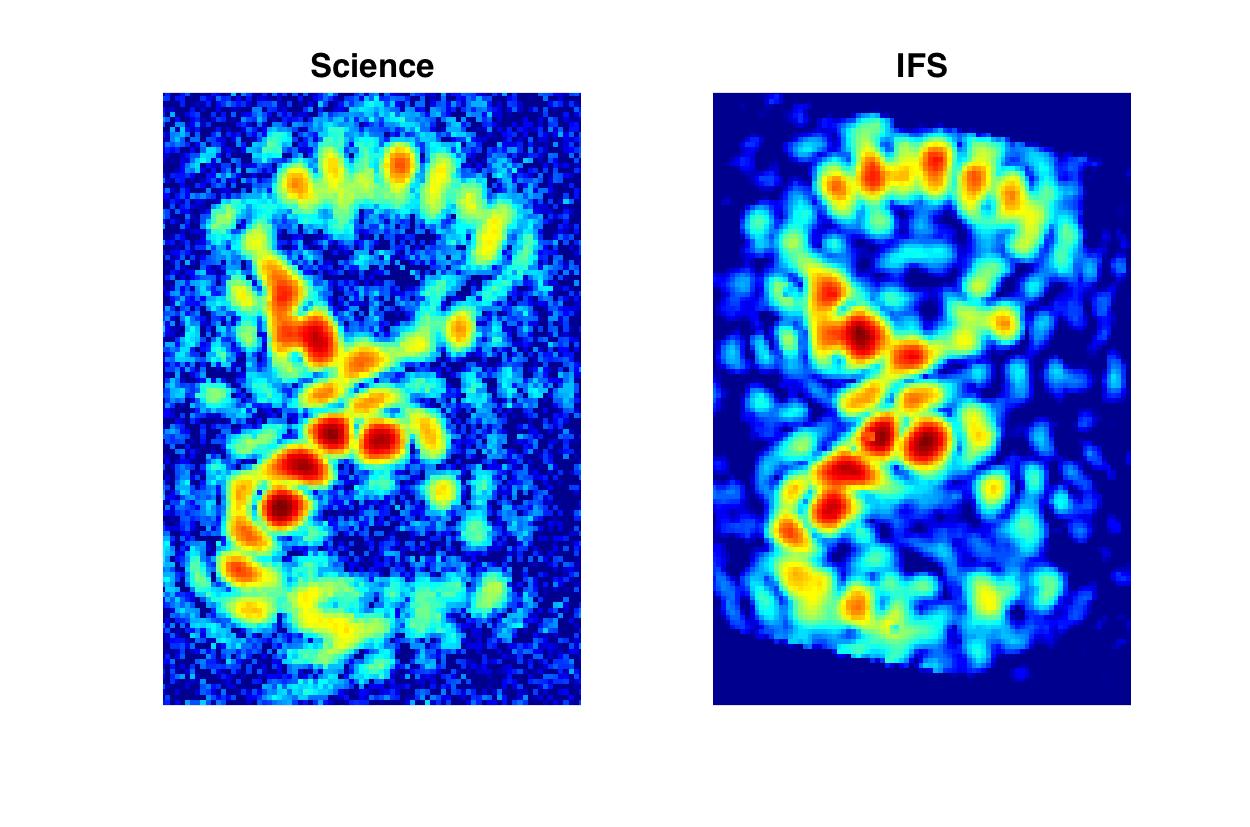}
\caption{A narrow band dark hole image on the science camera (left) versus the corresponding IFS slice (right).}
\label{fig:DHnb680}
\end{figure} 

Since the wavelength calibrations are produced using precisely chosen wavelengths and the PISCES bench has a stable gravity vector, lenslet PSFs should have sub-pixel stability.  Consequently, during the initial commissioning tasks, any arbitrary wavelength could have its position determined by interpolation using the calibration library. This enabled computationally fast and simple extraction, the production of a spectral covariance matrix and extraction that was optimal in the limit of Gaussian noise and well determined PSFs.  Extracted spectra was divided into 26 even slices in the three dimensional cube.  These slices correspond to the individual pixels on the detector that correspond to a full spectrum.  The extraction first determines positions of the border wavelengths of the 26 channels based on the centroid wavelength measurements obtained during the calibration.  Intensities for each channel are then obtained using a weighted sum over the pixels (including the use of fractional pixels).  The IFS reconstruction was compared to the science camera for a narrow band image in order to test the accuracy of the extraction.  This comparison, given in figure \ref{fig:DHnb680}, demonstrates the accuracy of the extraction given the similarity between the two images.

\subsection{Instrument Performance: Dark Hole Contrast} 
\label{sec:DarkHole} 

The commissioning goal for PISCES was to produce a narrow band dark hole contrast of on the order of 1e-8. Dark hole contrasts were measured using both narrow and broad band sources while manually adjusting the deformable mirrors.  PISCES satisfied this commissioning goal by a producing dark hole contrast of $\sim$1.4e-8 in a 30 second exposure at 680 nm.  This contrast was obtained after manually adjusting the DMs to achieve higher contrast.  The contrast ratio was determined by finding the peak to poke contrast on the detector and then comparing that to the averaged contrast ratio obtained in the regions given by the overlaid panda in figure 6.  Since, the average contrast ratio was used, the value was sensitive to the exact positioning of the panda over the different regions. Indeed, contrast ratios significantly better than 1e-8 are likely in local areas of the dark hole region.  Subsequent broadband testing post-commissioning demonstrated 1.1e-8 contrast in an 18\% band centered at 660nm, corresponding to WFIRST CGI IFS band $\#$1.  See the corresponding paper\cite{Groff2017} for more details.

\begin{figure}
\centering

\subfloat{\includegraphics[width=0.5\textwidth]{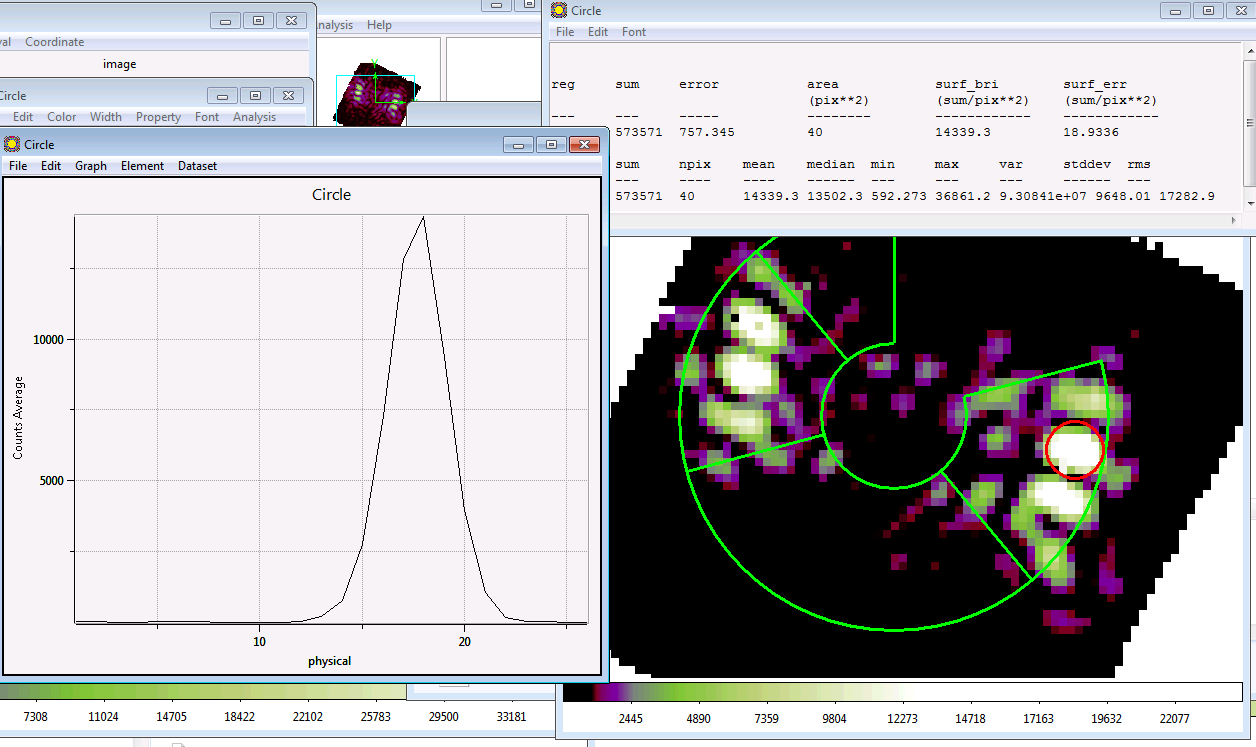}\label{fig:peakpoke}}
\subfloat{\includegraphics[width=0.5\textwidth]{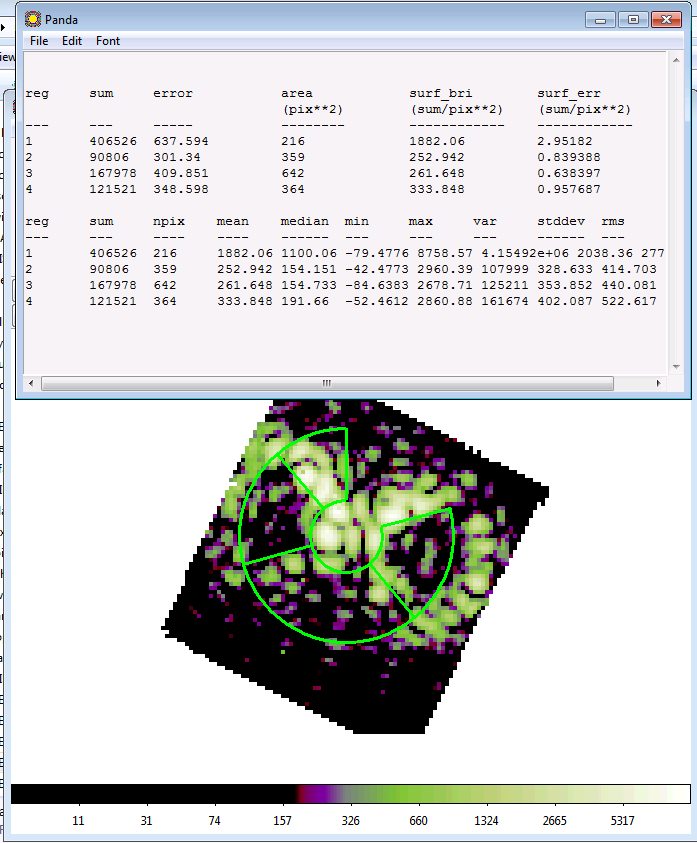}\label{fig:dhpanda}}

\caption{Images of the measurements used to satisfy the commissioning target: \protect\subref{fig:peakpoke} An image of the pokes used to determine peak to poke ratio. 
  \protect\subref{fig:dhpanda} The dark hole image with the panda used to determine average contrast overlaid.}

\end{figure}


 

\appendix    

\section*{Acknowledgements}
This research was carried out at Goddard Space Flight Center and the Jet Propulsion Laboratory, California Institute of Technology, under a contract with the National Aeronautics and Space Administration.
\bibliographystyle{spiebib}
\bibliography{biblibrary}

\end{document}